\documentclass[aps,pra,amsmath,amssymb,twocolumn,superscriptaddress]{revtex4}
\usepackage{amsmath,graphicx,float,epsfig,float,tabularx,multirow}
\usepackage{epstopdf,color,bbm,times}

\newcommand{\ket}[1]{| #1 \rangle}

\newcommand{\ketbra}[2]{| #1 \rangle \langle #2 |}

\def\imm{{\rm i}}

\begin{document}

\title{Homodyne-like detection for state-discrimination in the presence of phase noise}

\author{Matteo Bina}
\affiliation{Quantum Technology Lab, Department of Physics, University of Milan, via Celoria 16, I-20133 Milano, Italy}
\author{Alessia Allevi}
\affiliation{Department of Science and High Technology, University of Insubria and CNISM UdR Como, Via Valleggio 11, I-22100 Como, Italy}
\author{Maria Bondani}
\affiliation{Institute for Photonics and Nanotechnologies, CNR, and CNISM UdR Como, Via Valleggio 11, I-22100 Como, Italy}
\author{Stefano Olivares}
\email{stefano.olivares@fisica.unimi.it}
\affiliation{Quantum Technology Lab, Department of Physics, University of Milan, and INFN Sezione di Milano, via Celoria 16, I-20133 Milano, Italy}
\date{\today}

%
%
%
%
%

\begin{abstract}
We propose an innovative strategy to discriminate between two coherent states affected by either uniform or gaussian phase noise. The strategy is based on a homodyne-like detection scheme with photon-number-resolving detectors in the regime of low-intensity local oscillator. The experimental implementation of the detection scheme involves two hybrid photodetectors, whose outputs are used in post processing to calculate the shot-by-shot photon-number difference. The performance of this strategy is quantified in terms of the error probability in discriminating the noisy coherent signals as a function of the characteristic noise parameters.
\end{abstract}



\maketitle


{\it Introduction} - The present Letter aims at completing and enriching the research on optical communication schemes based on coherent states. Recently, many efforts have been devoted to find optimal discrimination strategies to minimize the detrimental effects of phase noise and to monitor the real-time communications by exploiting quantum estimation tools \cite{Muller,Becerra,Izumi}. In this respect, on the one hand it has been demonstrated that a homodyne detection scheme can implement a quasi-optimal discrimination strategy whenever a gaussian phase noise is present \cite{Oli:13}. On the other hand, we have already demonstrated that the employment of photon-number-resolving (PNR) detectors is crucial for the estimation of phase drifts in Kennedy-like receivers when only a single output port of the interferometer is monitored, even in the presence of phase noise \cite{BinaOlivares}. In this scenario, PNR detectors, providing direct access to the statistics of light, allow the gain of much more information than, e.g., single-photon detectors \cite{Sasaki_arxiv}.\\ 
To take advantage of the characteristics of both the standard homodyne detection scheme and PNR detectors, here we propose a hybrid detection scheme \cite{Wallentowitz,Banaszek,OLwigner,OE2012}, in which we use PNR detectors instead of pin-photodiodes and the difference between the two outputs of the interferometer is calculated in post processing. At variance with other existing homodyne-like detection schemes \cite{Puentes,Lahio,Zhang,Donati} , in our case the employment of hybrid photodetectors allows us to explore a wide photon-number dynamic range (up to 30 photons). For this reason, the detection apparatus, though being unable to detect optical states at the macroscopic level, is particularly useful to investigate different regimes of local oscillator (LO) intensity.\\ 
In this work we address a typical communication scheme with coherent signals, to which we apply the homodyne-like decision strategy. Namely, we evaluate the difference of photocounts and demonstrate that this reduces the error probability in discriminating the input coherent signals.
\par
{\it Homodyne detection with PNR detectors and state discrimination} - The ``imperfection'' in the discrimination protocol between non-orthogonal quantum states is quantified by the error probability $P_e$ and it depends on the employed measurement apparatus. Given two quantum states $\hat{\rho}_1$ and $\hat{\rho}_0$, with \textit{a priori} probabilities $\eta_1$ and $\eta_0$ satisfying $\eta_1+\eta_0=1$, the minimum error probability allowed by quantum mechanics is given by the Helstrom bound \cite{hel:76}
\begin{equation}\label{Helstrom}
P_e^{(H)}=\frac12 \Big [ 1-\text{Tr} |\eta_1 \hat{\rho}_1 -\eta_0 \hat{\rho}_0| \Big ].
\end{equation}
In our work we consider a standard interferometric scheme, in which a binary signal is encoded in two coherent states $\hat{\rho}_1\equiv\ketbra{\beta}{\beta}$ and $\hat{\rho}_0\equiv\ketbra{-\beta}{-\beta}$, namely the basic alphabet of a binary phase-shift-keyed (BPSK) communication. To be decoded, the state is then mixed at a beam splitter (BS) of transmittance $\tau$ with the LO $\ket{\alpha {\rm e}^{\imm \phi}}$. In our analysis we also assume that phase noise affects the propagation of the signals. The effect of a generic amount of phase noise on the coherent states may be described by the map
\begin{equation}\label{Noise}
\hat{\rho}_k\mapsto{\cal E}(\hat{\rho}_k)=\int_\mathbb{R} {\rm d}\varphi\, f(\varphi) \hat{U}(\varphi)\hat{\rho}_k \hat{U}^\dag(\varphi),
\end{equation}
where $\hat{U}(\varphi)=\exp(-\imm\varphi\hat{a}^\dag\hat{a})$ is the phase-shift operator, $[\hat{a},\hat{a}^\dag]=\hat{\mathbb{I}}$ and $f(\varphi)$ is a weight function describing the phase noise distribution. In a uniform phase noise scenario $f(\varphi)=\gamma^{-1} w(\varphi;\gamma)$, where $w(\varphi;\gamma)$ is a window function assuming the value of 1 inside the interval of phases $\varphi\in [-\gamma/2,\gamma/2]$ and 0 outside. In the case of gaussian phase noise, $f(\varphi)=\mathcal{N}(\varphi;\sigma^2)$ is the normal distribution of mean value 0 and variance $\sigma^2$.\\ In the presence of phase noise, by assuming $\eta_1=\eta_0=1/2$, the Helstrom bound (\ref{Helstrom}) is given by $P_e^{(H)}=\frac12 (1-{\rm Tr}|\hat{\Lambda}|)$, where $\hat{\Lambda}=\frac12 [{\cal E}(\hat{\rho}_1)-{\cal E}(\hat{\rho}_0)]$ or explicitly:
\begin{align}
\hat{\Lambda}=\int_\mathbb{R}{\rm d}&\varphi f(\varphi) {\rm e}^{-\beta^2}\nonumber\\
&\times\sum_{n,m}\frac{\beta^{n+m}{\rm e}^{\imm\varphi(n-m)}}{2\sqrt{n!m!}}\Big [ 1-(-1)^{n+m} \Big ] \ketbra{n}{m}.
\end{align}
\par
The evolved state at the two output ports of the BS is 
obtained by applying the unitary operator $\hat{U}_{\rm BS}=\exp\{\xi \hat{a}\hat{b}^\dag-\xi^*\hat{a}^\dag\hat{b}\}$, with $\hat{a}$ and $\hat{b}$ describing the input modes and $\xi$ characterizing the BS transmissivity $\tau=\cos^2\xi$. If the LO is a high-intensity coherent state, homodyne detection in the presence of phase noise results to be a quasi-optimal discrimination strategy \cite{Oli:13} with an overall error probability given by
\begin{equation}\label{PeHD}
P_e^{({\rm hd})}=\frac12 \left [ \int_{-\infty}^0 p(x;\beta) + \int^{+\infty}_0 p(x;-\beta) \right ],
\end{equation}
where
\begin{equation}
p(x;\pm\beta)=\frac{1}{\sqrt{\pi}}\int_\mathbb{R}{\rm d}\varphi f(\varphi){\rm e}^{-(x\,\mp\sqrt{2}\,\beta\cos\varphi)^2}.
\end{equation}
\par
In our scheme we consider low-intensity coherent states for both signal and LO. Now, the output modes of the BS, which we label $\hat{c}$ and $\hat{d}$ hereafter, can be monitored by means of PNR detectors, giving access to the photon statistics of the output states. At the same time, the difference of the measured discrete photo-currents is equivalent to a homodyne-like detection, which measures the field quadratures of the input signals. The distribution of the aleatory variable $y=n-m$, with the two stochastic variables $n$ and $m$ described by  Poisson distributions having mean values $\mu_n$ and $\mu_m$, is given by the Skellam distribution \cite{skellam}
\begin{equation}\label{Skellam}
S_y(\mu_c,\mu_d)={\rm e }^{-\mu_c-\mu_d}\left ( \frac{\mu_c}{\mu_d} \right )^{\frac{y}{2}}I_y(2\sqrt{\mu_c\mu_d})\qquad y\in \mathbb{Z},
\end{equation}
where $I_y(x)$ is the modified Bessel function of the first kind. In the ideal case of the absence of phase noise, the mean values of the Poisson distributions are
\begin{subequations}\label{PoissonAve}\begin{align}
\mu_{\pm,c}&=\alpha^2(1-\tau)+\beta^2\tau\pm 2\sqrt{\tau(1-\tau)}\alpha\beta\cos\phi \\
\mu_{\pm,d}&=\alpha^2\tau+\beta^2(1-\tau)\mp 2\sqrt{\tau(1-\tau)}\alpha\beta\cos\phi,
\end{align}\end{subequations}
depending on which coherent state $\ket{\pm\beta}$ of the BPSK alphabet has been sent. To fix the ideas, let us assume that $\mu_{+,c}>\mu_{+,d}$ and $\mu_{-,c}<\mu_{-,d}$. Then we expect to detect a higher (lower) number of photons in $\hat{c}$ than in $\hat{d}$ if $\ket{\beta}$ ($\ket{-\beta}$) has been sent. Whenever the photon detection does not satisfy these expectations, an error in the inference about the sent signal occurs.\\
We define the photocount differences $\Delta_1\equiv\langle \hat{n}_c-\hat{n}_d\rangle_1$ and $\Delta_0\equiv\langle \hat{n}_c-\hat{n}_d\rangle_0$, depending on which coherent signal $\ket{\pm\beta}$ has been sent. We point out that $\hat{n}_c=\ketbra{n_c}{n_c}$ and $\hat{n}_d=\ketbra{n_d}{n_d}$ are the single-shot measurements of the PNR detectors at the output modes, so that the photocount differences are integer numbers $\{\Delta_0,\Delta_1\} \in \mathbb{Z}$. Thus, the expected overall error probability in the discrimination process can be readily calculated for the Skellam distribution and it reads as
\begin{equation}\label{PeSkellam}
p_e^{(sk)}(\phi)= \sum_{\Delta_1=-\infty}^{-1} S_{\Delta_1}\big (\mu_{+,c},\mu_{+,d}\big ) +\frac12 S_0,
\end{equation}
where $S_{\Delta_1}\big (\mu_{+,c},\mu_{+,d}\big ) = S_{\Delta_0}\big (\mu_{-,c},\mu_{-,d}\big ) $ and $S_0$ is the value of the Skellam distribution for $\Delta_1=\Delta_0=0$, i.e. in the case of inconclusive measurement.\\
If we take into account phase noise, the mean numbers of photons at the outputs of the BS are given by Eqs. (\ref{PoissonAve}) with the substitution $\phi\to\phi-\varphi$. Thus, the overall error probability is obtained by integrating Eq. (\ref{PeSkellam}) with the weight function describing the corresponding noise model:
\begin{equation}\label{PeSkellamNoise}
P_e^{(sk)}=\int_\mathbb{R}{\rm d}\varphi f(\varphi) p_e^{(sk)}(\phi-\varphi).
\end{equation}
\par
\begin{figure}[b!]
\centering
\includegraphics[width=.85\linewidth]{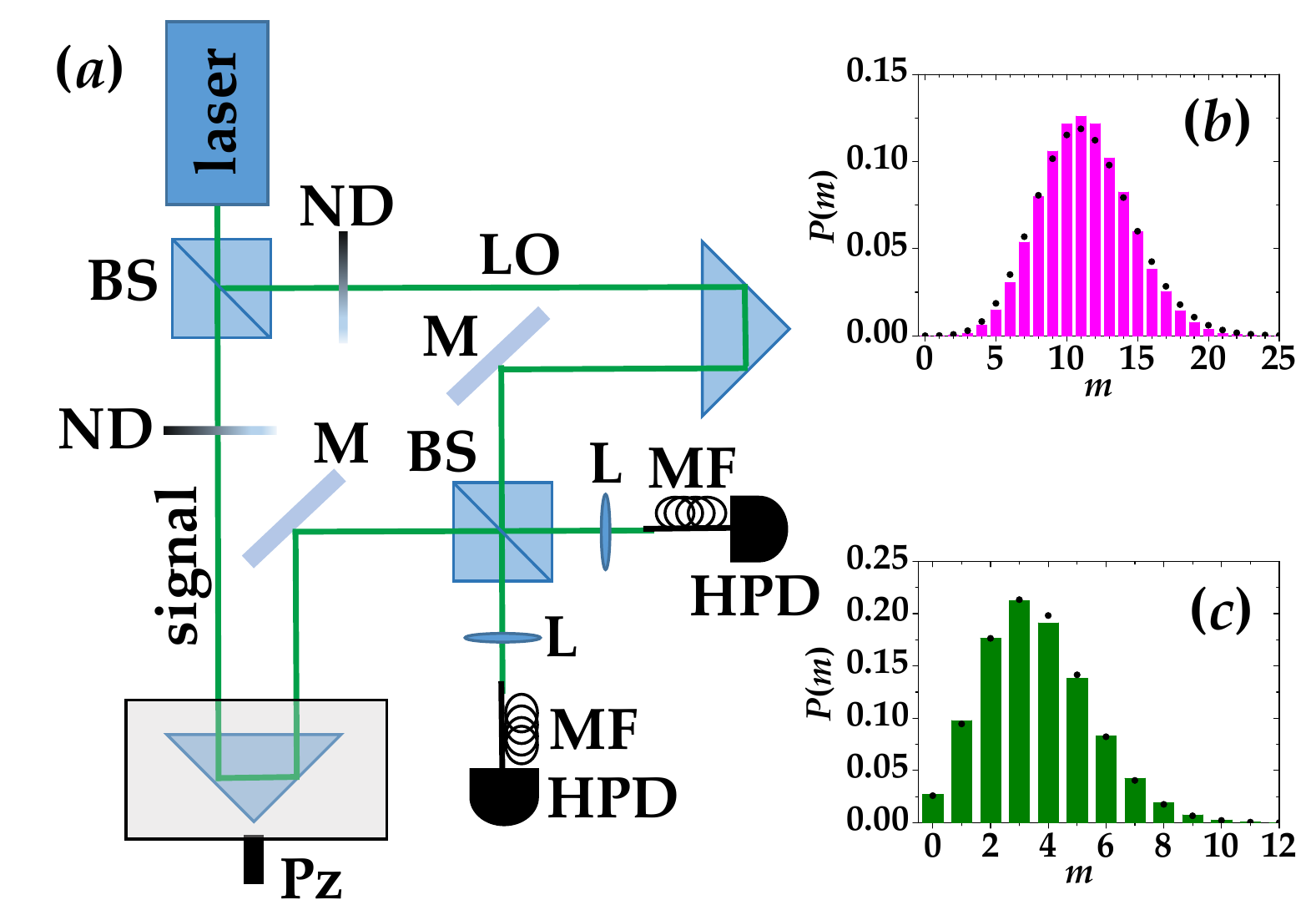}
\vspace{0 cm}
\caption{(a) Sketch of the experimental setup, in which BS is a balanced beam splitter, M is a high-reflectance mirror, Pz is the piezoelectric movement, ND is a variable neutral density filter, L is a lens, MF is a multi-mode fiber, HPD is the hybrid photodetector. (b) and (c): Two typical photon-number distributions measured by the HPDs. Columns: Experimental data, black dots: Theoretical Poisson distribution with the same mean value as the data.}
\label{f:setup}
\end{figure}

{\it Proof-of-principle experiment} - In order to test the performance of our strategy, i.e. the employment of PNR detectors in a homodyne-like measurement with a low-intensity LO and in the presence of phase noise, we realized a proof-of-principle experiment.\\
As shown in Fig.~\ref{f:setup}, the second-harmonic pulses (5-ps-pulse duration) emitted at 523 nm by a mode-locked Nd:YLF laser regeneratively amplified at 500 Hz were sent to a Mach-Zehnder interferometer to get the signal and the LO. In order to change the balancing between the two fields, we inserted two variable neutral density filters in the two arms. In particular, we chose two different configurations: In the first one we mixed a coherent signal with a LO of similar amplitude, whereas in the second one we introduced a significant unbalancing between the two amplitudes. In both situations, we optimized the spatial and temporal superposition of signal and LO in order to get almost the best overlap admitted by the choice of the amplitudes and of the balancing.\\ 
The length of one arm of the interferometer was changed in steps by means of a piezoelectric movement in order to modify the LO phase in the whole $2\pi-$range. The light at the two outputs of the second BS was collected by two multi-mode fibers (600-$\mu$m-core diameter) and sent to two hybrid photodetectors (HPD, mod. R10467U-40, Hamamatsu). The output of each detector was amplified (preamplifier A250 plus amplifier A275, Amptek), synchronously integrated over a 500-ns window (SGI, SR250, Stanford) and digitized (AT-MIO-16E-1, National Instruments). We set 60 different piezo positions and for each one we saved 50000 laser shots. Typical reconstructions of the photon-number statistics registered by each detector are shown in Fig.~\ref{f:setup}, from which it is possible to appreciate the wide dynamic range of HPDs. We notice that in both panels (b) and (c) theoretical Poisson distributions are well superimposed to the experimental data, thus confirming the correctness of the model expressed by Eqs.~(\ref{PoissonAve}).
As already explained in \cite{josaB10,BinaOlivares}, by exploiting the linearity of HPDs it is possible to extract information about the phase. In fact, the mean number of photons detected at each output of the interferometer describes the interference pattern as a function of the piezo position. Thus, the relative phase between the two arms of the interferometer can be retrieved by normalizing the mean values between -1 and 1, and by applying the arcos function. We also notice that, from the mean number of photons detected at each output as a function of the phase, it is possible to extract the effective portions of LO ($a_c$ and $a_d$) and signal ($b_c$ and $b_d$) either transmitted or reflected by the BS. Indeed, the mean values at the two outputs are linked to these quantities via $\mu_{c}=a^2_c+b^2_c + 2 a_c b_c\cos(\phi)$ and $\mu_{d}=a^2_d+b^2_d + 2 a_d b_d\cos(\phi)$. By using these expressions, we got $a_c = 2.01$, $a_d = 2.07$, $b_c = 1.13$, and $b_d = 1.07$ in the more balanced case (Experiment \#1), and $a_c = 2.74$, $a_d = 2.68$, $b_c = 0.87$, and $b_d = 0.85$ in the less balanced one (Experiment \#2). The monitoring of the mean values also allowed us to check the stability of the detection apparatus during the long measurement sessions.\\ 
\begin{figure}[t]
\centering
\includegraphics[width=0.49\linewidth]{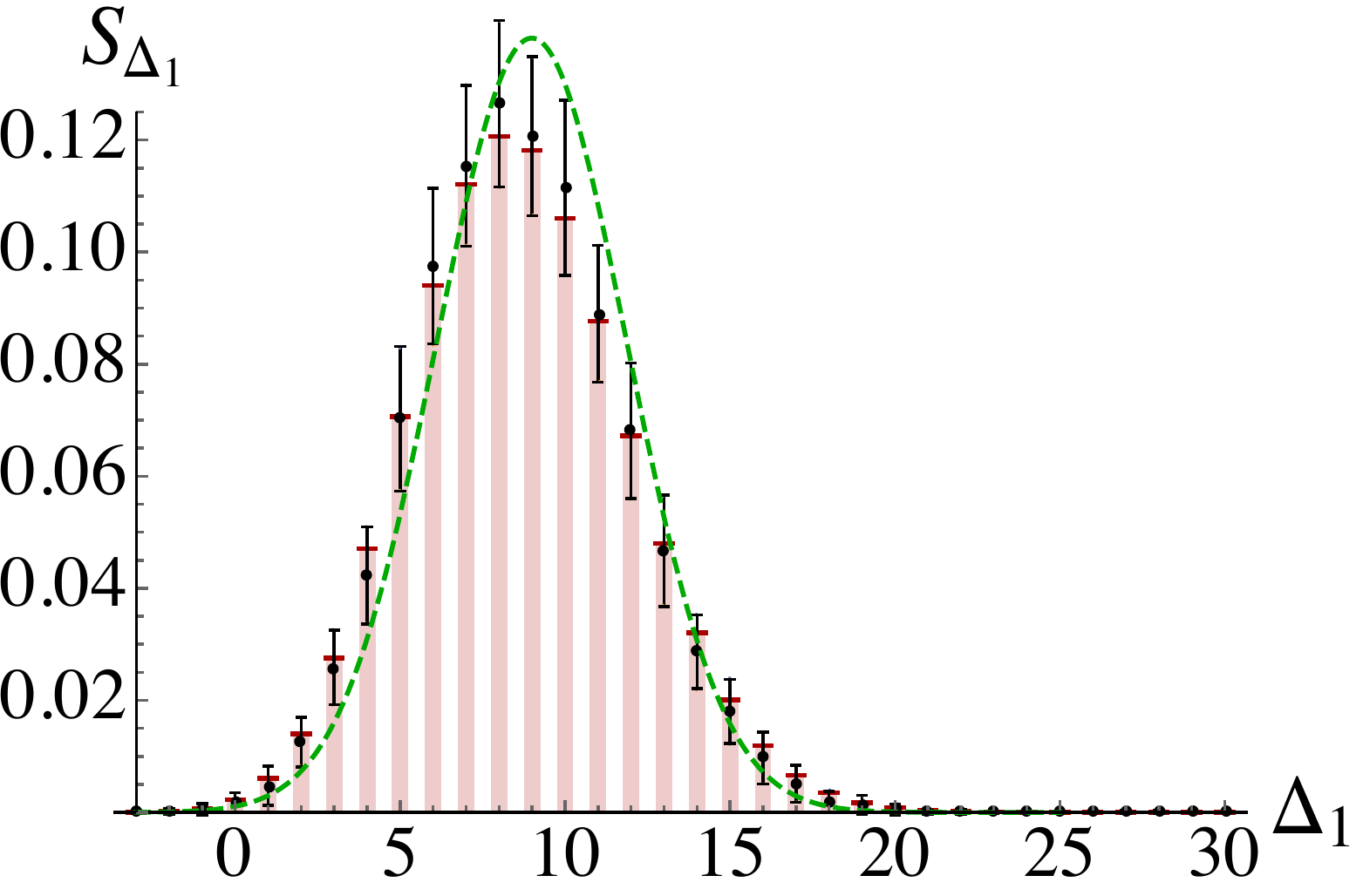}
\includegraphics[width=0.49\linewidth]{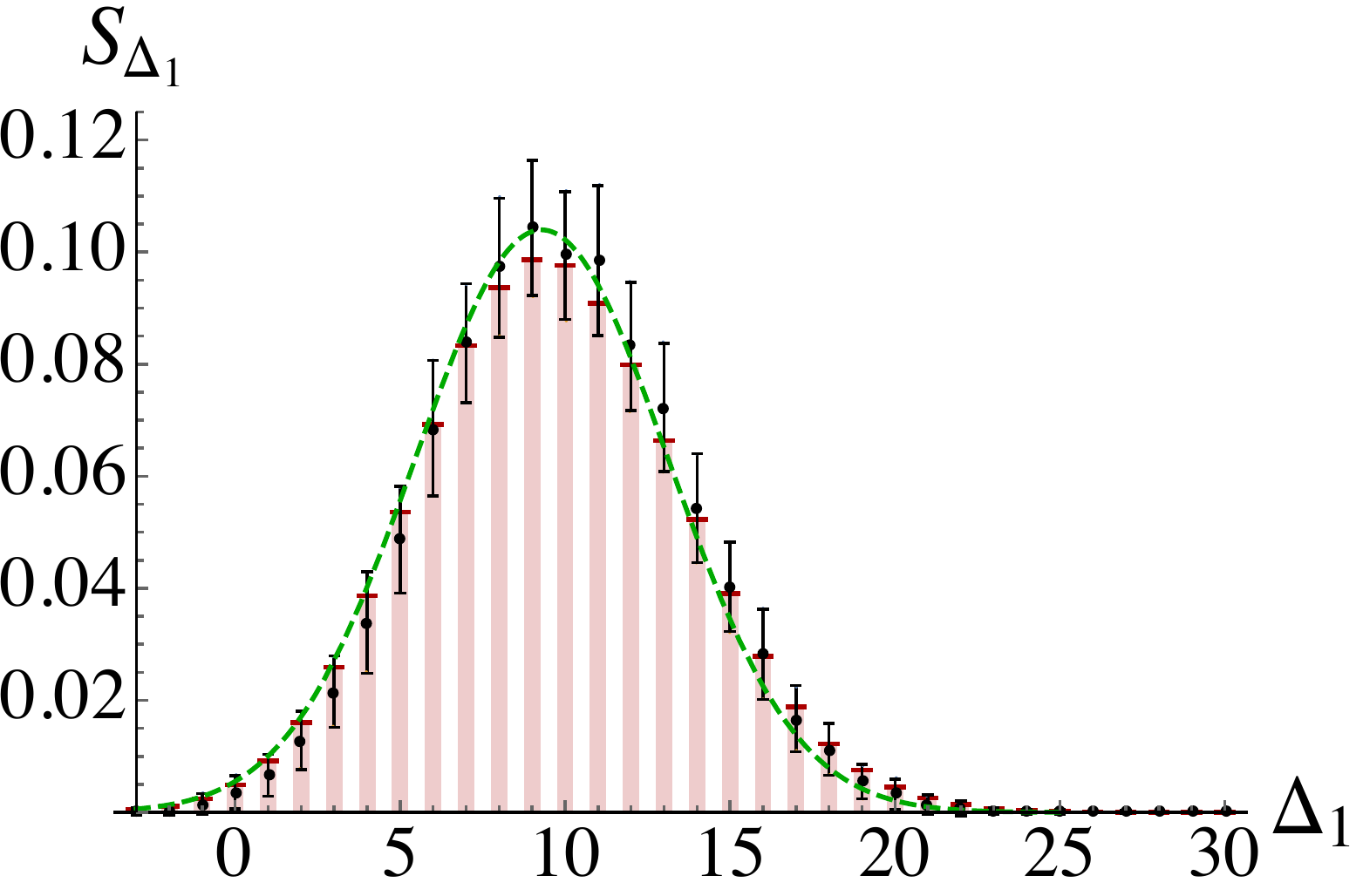}
\caption{Plots of the probability distribution of the experimental photocount differences $\Delta_1$, for both Experiment \#1 (left panel) and Experiment \#2 (right panel), when the relative phase between signal and LO is $\phi=0$. The experimental data, averaged over $M=100$ repetitions for bootstrapping (dots), are well fitted by the histogram plot representing the theoretical Skellam distribution (\ref{Skellam}). The agreement between homodyne probability distribution (green dashed) and experimental data is better in the presence of a significant unbalancing between signal and LO (right panel).}
\label{f:skellam}
\end{figure}
Once we have assigned a phase value to each piezo position, the two reference coherent states $\ket{\pm\beta}$ can be obtained by choosing the LO phase $\phi=0$ and $\phi=\pi$, respectively. Moreover, proper sets of data samples can be combined together to generate the mixtures of coherent states affected by a different amount of either uniform or gaussian phase noise. Given one of these sets of data, for each output mode we randomly selected $N_{\rm s}=10^3$ photocounts, which were used to obtain the shot-by-shot photon-number differences. Then, for each choice of the input state, we evaluated the error probability in the state discrimination by normalizing the number of wrong values of the photon-number difference to the total number $N_s$ of selected photocounts. We repeated such an operation $M=100$ times by applying a bootstrapping procedure \cite{JCGM}. This post-processing of data constitutes the heart of the proof-of-principle experiment. It allows us to obtain the expected error probability and to compare the results with the theoretical expectations for the standard homodyne measurement and the ultimate quantum limit set by the Helstrom bound.
In Fig.~\ref{f:skellam}, we show the plots of the probability distribution of the photocount differences obtained with Experiment \#1 (left panel) and Experiment \#2 (right panel). The relative phase between signal and LO has been set to $\phi=0$ and the experimental points represent the average over $M=100$ repetitions employed for the bootstrap procedure. In both cases, the theoretical expectations given by the Skellam distributions and calculated according to Eq.~(\ref{Skellam}), evaluated at the experimental values of $\mu_{c}$ and $\mu_{d}$, are well-superimposed to the experimental data. In the same figure we also show the corresponding homodyne distribution concerning coherent-state discrimination, which appears, as expected, closer to the Skellam in the less balanced case.  
\par
\begin{figure}[t!]
\centering
\includegraphics[width=0.49\linewidth]{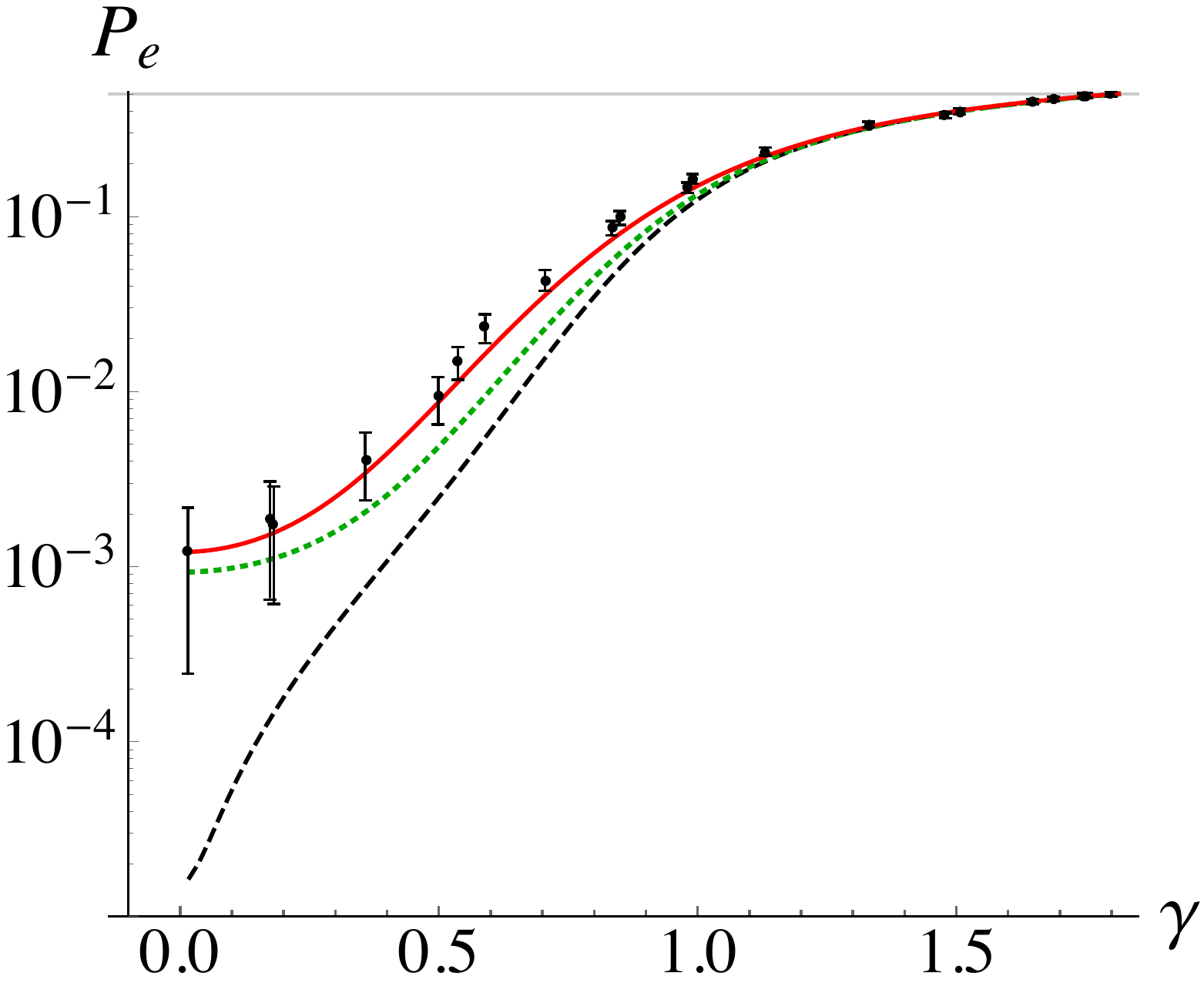}
\includegraphics[width=0.49\linewidth]{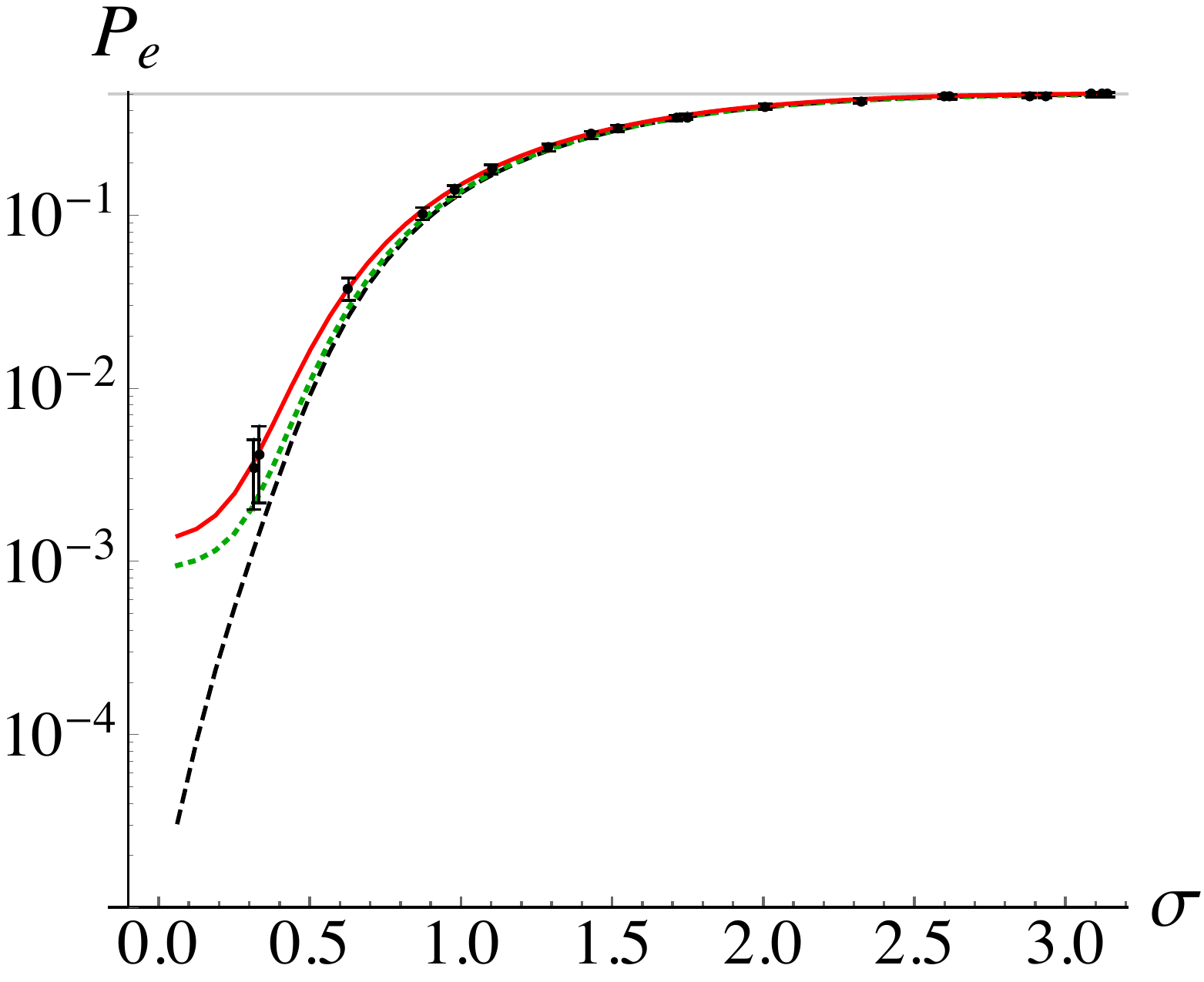}
\caption{Experiment \#1 (signals and LO with similar intensities): Error probability (in logarithmic scale) obtained by the experimental homodyne-like photocount differences (black dots with error bars) as a function of the uniform phase noise parameter (left panel) and gaussian phase noise one (right panel). In each panel we also plot the theoretical prediction using the Skellam distribution (solid red curve), the corresponding theoretical standard homodyne detection (green dotted curve) and the Helstrom bound (black dashed curve). }
\label{f:experiment1}
\end{figure}
\begin{figure}[t!]
\centering
\includegraphics[width=0.49\linewidth]{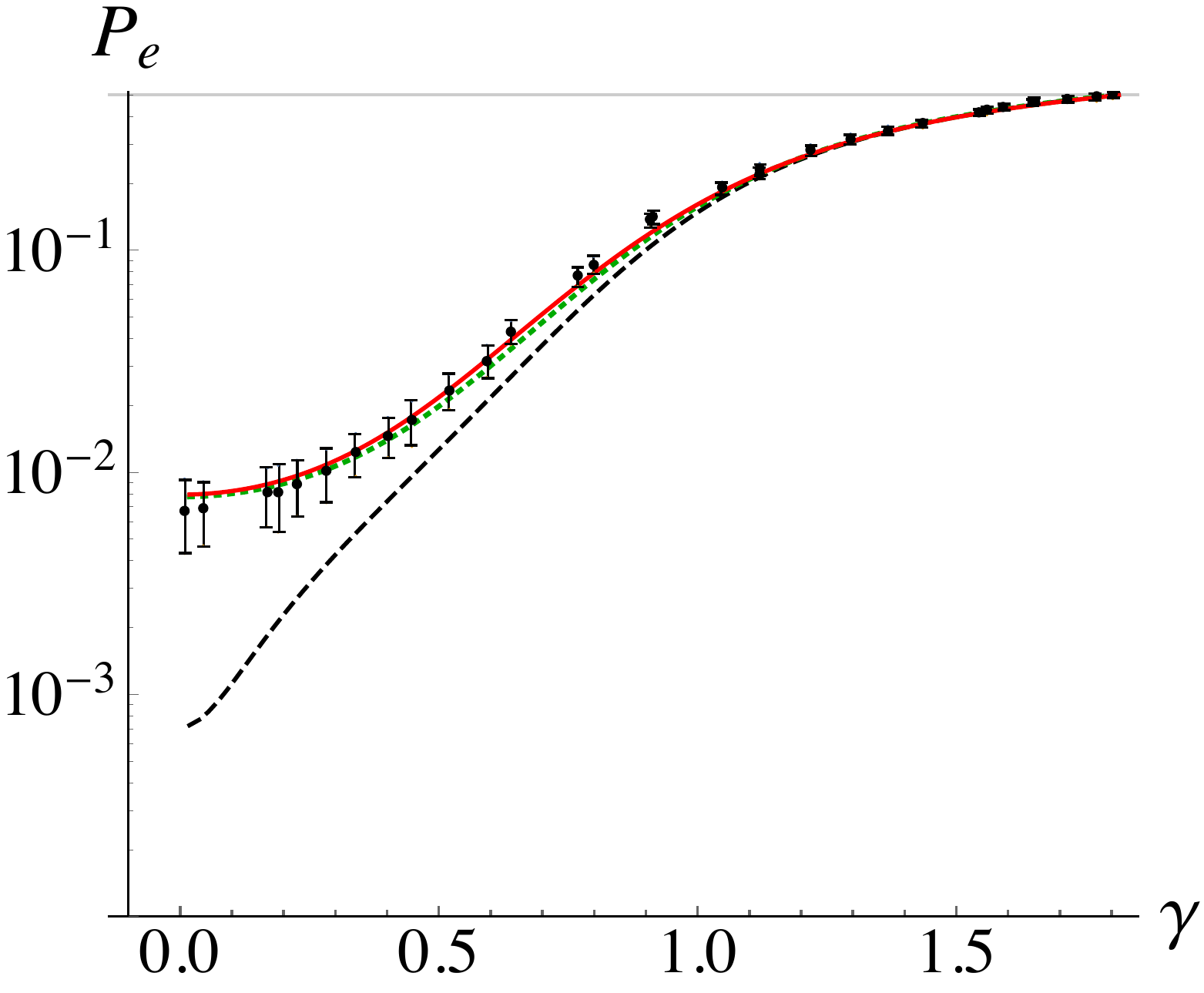}
\includegraphics[width=0.49\linewidth]{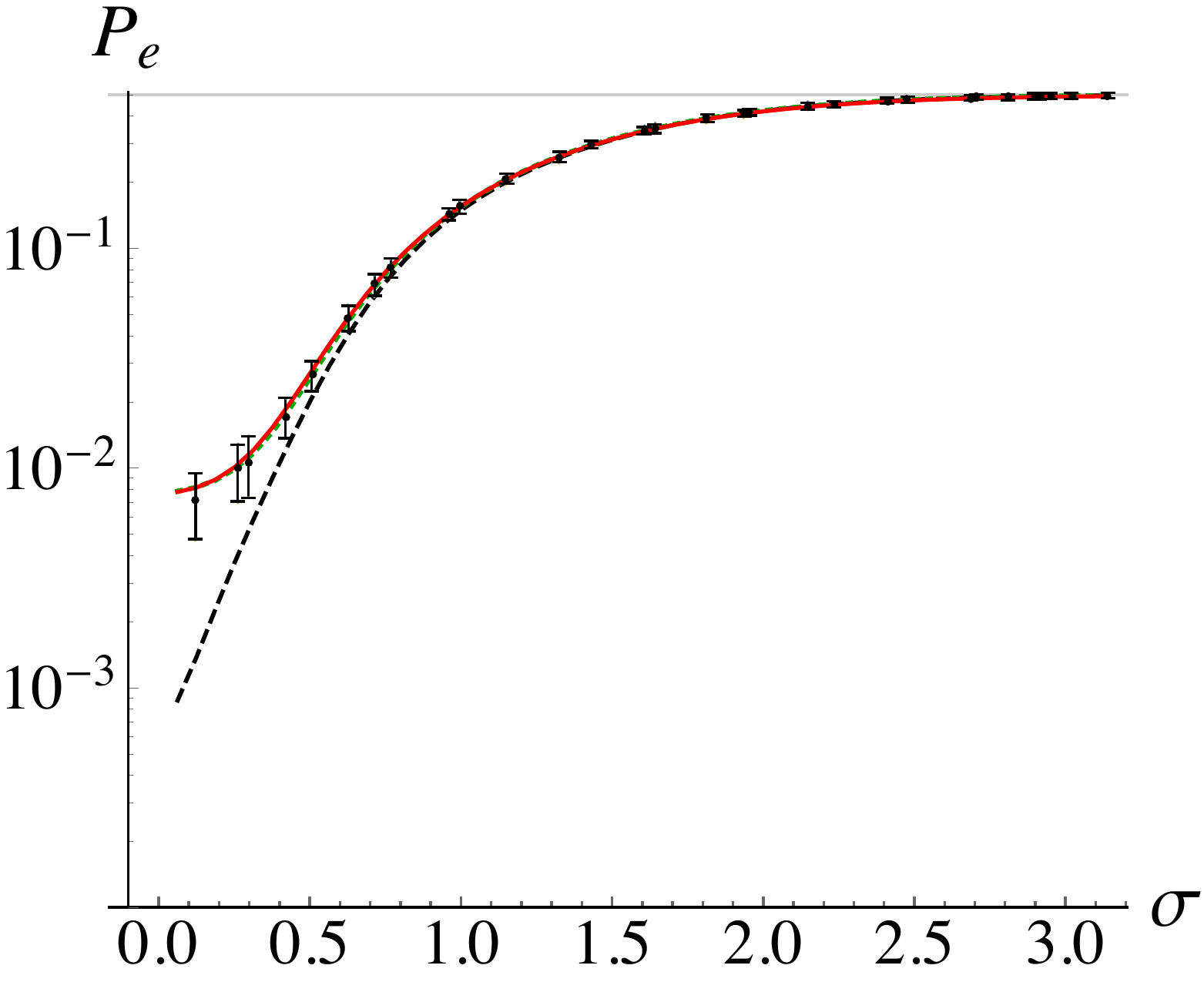}
\caption{Experiment \#2 (signals and LO with different intensities):  the description of these plots is the same as in Fig. \ref{f:experiment1}. }
\label{f:experiment2}
\end{figure}

%
%
{\it Results} - The experimental results are shown in Figs. (\ref{f:experiment1}) and (\ref{f:experiment2}) for the two configurations. The experimental error probability distributions plotted in the two figures have been obtained for different values of the uniform noise parameter $\gamma$, and the gaussian noise standard deviation $\sigma$. 
In both figures, the error bars, corresponding to the experimental data, have been obtained by applying the bootstrap statistical method.  
We note that the experimental results remarkably agree with the theoretical prediction (solid red curve) given by Eq. (\ref{PeSkellamNoise}). 
\par
These results are compared with the standard homodyne technique (corresponding to the green curve in the figures), in which a high-intesity LO is employed. We notice that such a curve represents the theoretical expectation evaluated for the two experimental choices of the signal field amplitudes.
We observe that, in the case in which signals and LO have more similar amplitudes (Experiment \#1), the homodyne-like measurement with PNR detectors is very close to the standard homodyne technique (Fig. \ref{f:experiment1}). 
\begin{figure}[t!]
\centering
\includegraphics[width=0.51\linewidth]{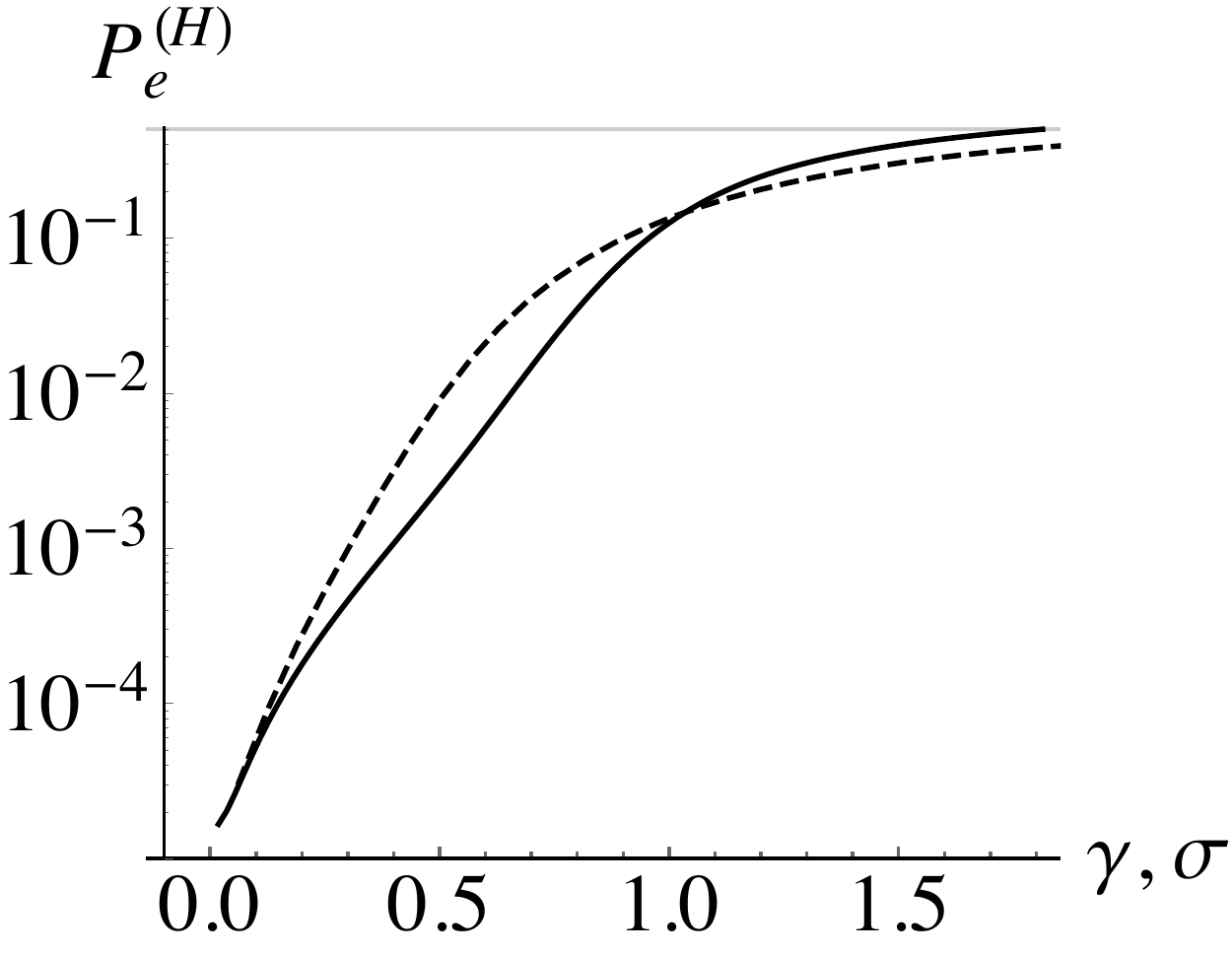}
\includegraphics[width=0.48\linewidth]{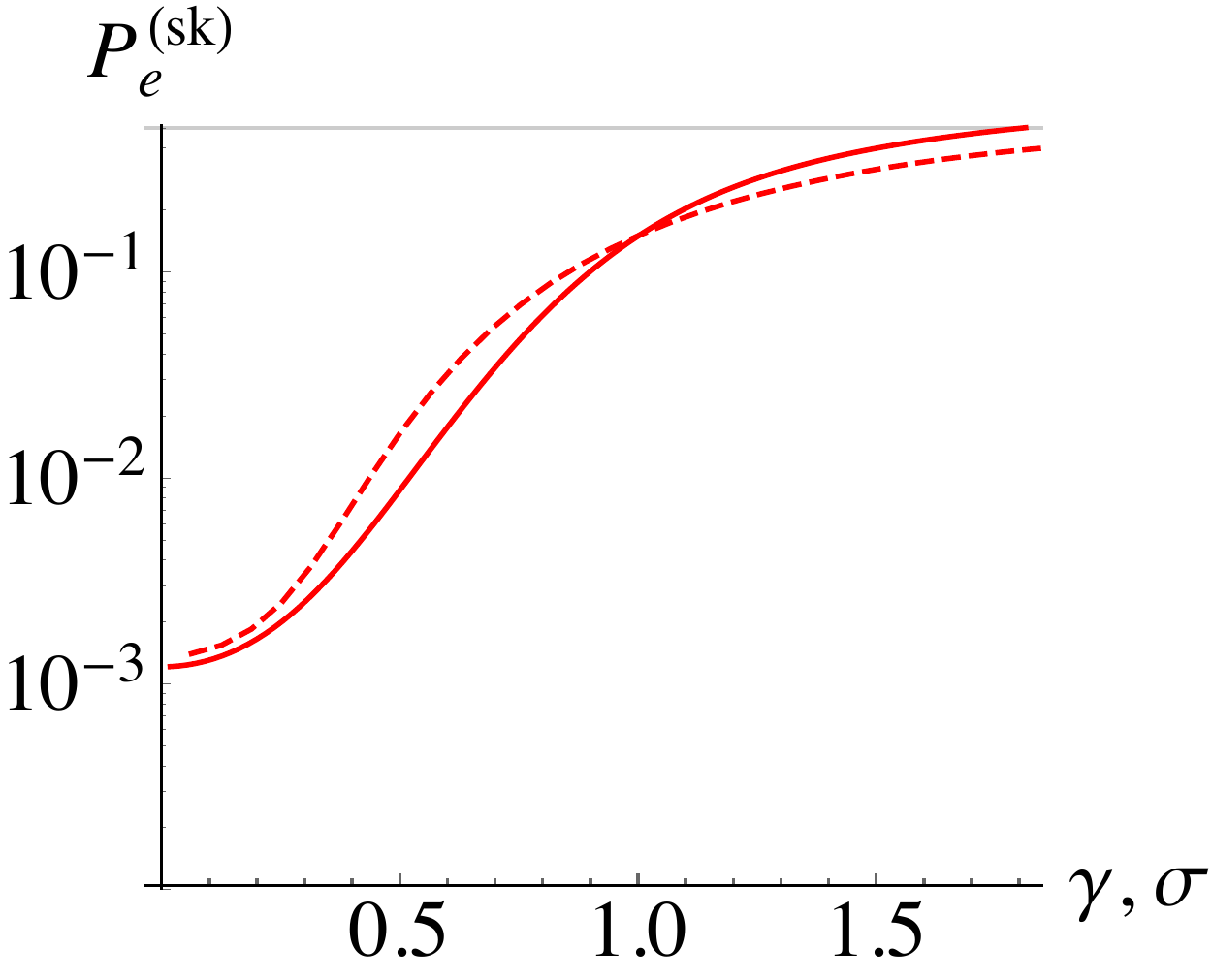}
\caption{Comparison between uniform (solid curves) and gaussian (dashed curves) noise for the error probabilities $P_e^{(H)}$ and $P_e^{(sk)}$ obtained from the Helstrom bound (left panel) and the photocounts difference (right panel), respectively.}
\label{f:uniform_gaussian}
\end{figure}
In Experiment \#2, where signal and LO amplitudes are significantly different, the two techniques provide almost the same results (see Fig. \ref{f:experiment2}), which tend to become more similar by increasing the LO intensity. In such a way the system resembles the typical homodyne scheme. In both cases, the homodyne-like measurements with PNR detectors, as well as the standard homodyne technique, result to be quasi-optimal for state discrimination. In fact, even for moderately small values of the noise parameters, the red and green curves in Figs.~\ref{f:experiment1} and ~\ref{f:experiment2} approach the black one, corresponding to the theoretical expectation of the Helstrom bound for the experimental choices of the signal field amplitudes. This convergence demonstrates that not only the standard homodyne scheme, but also our homodyne-like detection scheme based on PNR detectors is quasi-optimal for coherent-state discrimination in the presence of phase noise.\\
It is worth noting that the obtained error probabilities display a different behavior when the two phase noise models are employed. In order to make a fair comparison between them, we equate the variances of the two phase noise distributions, thus obtaining the relationship $\sigma=\gamma/(2\sqrt{3})$ between the two noise parameters.
From Fig. \ref{f:uniform_gaussian}, it is evident that, for some values of the uniform noise parameter $\gamma$, the error probabilities corresponding to Eqs. (\ref{Helstrom})-(\ref{PeSkellam}), are below those obtained for the gaussian-noise case. 
\par
{\it Conclusions} - We presented the implementation of a homodyne-like detection scheme based on HPDs in the regime of low-intensity signals and LO. Such a scheme has been used to investigate, both theoretically and experimentally, the problem of coherent-state discrimination in BPSK communication. In particular, we experimentally demonstrated that the efficiency of our detection technique in addressing the shot-by-shot discrimination protocol is very similar to, and in some cases indistinguishable from, that of the standard homodyne technique employing a high-intensity LO. We characterized this discrimination strategy for both uniform and gaussian phase noise, showing that there are threshold values of the noise parameters for which the error probability is minimal. This result testifies that our strategy, besides the standard homodyne technique, is quasi-optimal for coherent-state discrimination in the presence of phase noise. We believe that our proposal is well suited to those quantum communication schemes employing low-intensity coherent states and PNR detectors, mainly in the presence of phase noise. Finally, the hybrid scheme we realized, giving us direct access to the photon-number distributions at each output of the interferometer, could be used to implement a continuous-variable cryptographic scheme with nonclassical states, such as squeezed and sub-Poissonian states.\\

{\it Acknowledgments} - SO thanks Matteo Paris for stimulating and useful discussions and Luca Sguera for his support in the early stage of this work.

%


\end{document}